\documentclass[sigconf]{acmart} %
\AtBeginDocument{%
  }

\copyrightyear{2026}
\acmYear{2026}
\setcopyright{cc}
\setcctype{by-nc-nd}
\acmConference[UIST '26]{The 39th Annual ACM Symposium on User Interface Software and Technology}{November 02--05, 2026}{Detroit, MI, USA}
\acmBooktitle{The 39th Annual ACM Symposium on User Interface Software and Technology (UIST '26), November 02--05, 2026, Detroit, MI, USA}
\acmDOI{10.1145/3830398.3830489}
\acmISBN{979-8-4007-2856-3/2026/11}

\usepackage{siunitx}            %
\usepackage{xspace}             %
\newcommand{\thermosaic}{\textsc{TherMosaic}\xspace}

\begin{document}

\title[\thermosaic]{\thermosaic: Accelerating Perceived Thermal Transitions Through Spatiotemporal Thermal Feedback}

\author{Zining Zhang}
\email{znzhang@umd.edu}
\affiliation{%
  \institution{University of Maryland}
  \city{College Park}
  \state{Maryland}
  \country{USA}}

\author{Jiasheng Li}
\email{jsli@umd.edu}
\affiliation{%
  \institution{University of Maryland}
  \city{College Park}
  \state{Maryland}
  \country{USA}}

\author{Myungin Lee}
\email{myungin@umd.edu}
\affiliation{%
  \institution{University of Maryland}
  \city{College Park}
  \state{Maryland}
  \country{USA}}

\author{Zeyu Yan}
\email{zeyuy@umd.edu}
\affiliation{%
  \institution{University of Maryland}
  \city{College Park}
  \state{Maryland}
  \country{USA}}

\author{Jin Ryong Kim}
\email{jin.kim@utdallas.edu}
\affiliation{%
  \institution{University of Texas at Dallas}
  \city{Richardson}
  \state{Texas}
  \country{USA}}

\author{Huaishu Peng}
\email{huaishu@umd.edu}
\affiliation{%
  \institution{University of Maryland}
  \city{College Park}
  \state{Maryland}
  \country{USA}}

\renewcommand{\shortauthors}{Zhang et al.}

\begin{abstract}
Thermal feedback can enrich immersive interaction, but thermoelectric devices often change temperature too slowly to match interactive timing. We present \thermosaic, a spatiotemporal thermal feedback approach that accelerates perceived temperature transitions by leveraging two perceptual mechanisms: spatial summation and thermal adaptation. Focusing on the fingertip, we first investigate this approach using a custom 2$\times$2 array of independently controlled Peltier modules. Across three controlled perceptual studies, we show that distributed thermal stimulation can preserve stable hot and cold percepts despite local deviations, that adaptation helps maintain these percepts during changing stimulation, and that combining these effects reduces perceived transition time by about 30\%--40\% for transitions originating from hot or cold states.  We then translate the same design principles into a standalone wearable implementation of \thermosaic and evaluate it in virtual reality. Our results show that this approach reduces perceived thermal lag and improves temporal alignment between thermal and visual events in interactive use.
\end{abstract}


\begin{teaserfigure}
  \includegraphics[width=\textwidth]{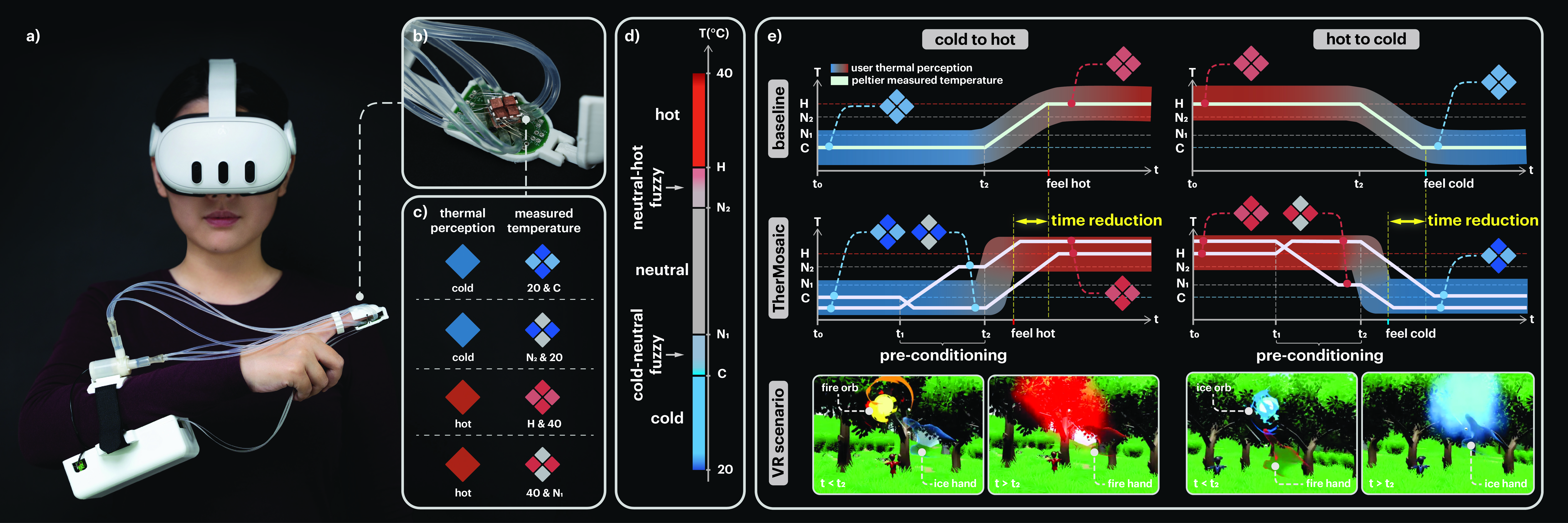}
  \caption{
  (a) \thermosaic standalone wearable implementation.
  (b) Fingertip thermal display.
  (c) Examples of reported user thermal perception alongside corresponding per-Peltier measured temperatures.
  (d) Color-coded temperature scale showing perceptual categories and their boundaries.
  (e) Comparison of per-Peltier measured temperature changes over time and user thermal sensation between the baseline condition (Peltier producing uniform temperature) and \thermosaic. 
  Our approach preconditions partial thermal inputs to maintain the current percept while reducing the apparent perceptual distance between categories, accelerating perception of transitions (e.g., switching between fire and ice magic in VR).}
  \Description{A multi-panel figure illustrating a wearable fingertip thermal display system designed for virtual reality applications.
  (a) A standalone wearable implementation of the TherMosaic device, shown mounted on a user’s hand, with integrated electronics and multiple thermal actuators. 
  (b) A close-up view of the fingertip thermal display, highlighting an array of small Peltier elements arranged to deliver localized temperature stimuli to the skin.  
  (c) A set of example illustrations showing reported user thermal perceptions (e.g., cold, neutral, warm, hot) alongside the corresponding temperatures measured at each Peltier element, demonstrating how physical temperatures map to subjective sensations.
  (d) A color-coded temperature scale that categorizes perceptual ranges with marked boundaries between categories. 
  (e) A comparison graph showing temperature changes over time at individual Peltier elements alongside user-reported sensations, contrasting a baseline condition with the TherMosaic approach, and further illustrated through example VR scenarios involving fire and ice magic interactions.}
  \label{fig:teaser}
\end{teaserfigure}

\maketitle

\section{Introduction}\label{intro}
Thermal feedback has long been envisioned as a powerful channel for enhancing immersion, realism, and emotional expression in interactive systems and virtual reality. It communicates material qualities~\cite{material_discrimination, yang2009spatial}, evokes affective cues~\cite{multi_moji, thermal_taste, emotion_image, emotion_vib, thermon, thermo_photo}, and enriches multisensory experiences~\cite{thermal_taste, LYU2023102279, da2023thermal, jeong2025interactive}. 
Yet despite this promise, thermal feedback remains difficult to integrate into interactive settings because temperature transitions are often too slow to match the timing of user actions and virtual events. 
Most existing systems rely on thermoelectric actuators such as Peltier modules, whose heating and cooling dynamics are inherently gradual and often require several seconds or longer to switch between thermal states~\cite{choi2021lightweight}. This lag can cause thermal sensations to fall out of synchrony with interaction events, weakening perceptual continuity and diminishing realism. 

Prior approaches have largely treated this limitation as a hardware problem. Researchers have therefore sought to accelerate heat transfer through hardware innovations such as ultra-thin substrates~\cite{thin_film}, high-conductivity materials~\cite{high_conductivity_AIN}, and liquid-cooling mechanisms~\cite{cooling_fluids}. While effective, these approaches often increase system complexity, size, or power demands, which can limit their suitability for compact interfaces such as fingertip displays. 

In this paper, we take a different perspective. Rather than seeking to increase the physical temperature ramp rate through hardware improvements, we ask whether thermal transitions can instead be made to \textit{feel} faster. We propose \thermosaic, a spatiotemporal thermal feedback approach that accelerates perceived temperature transitions on the fingertip by coordinating multiple thermal subregions. We focus on the fingertip because it is both thermally sensitive and central to touch-based interaction, yet its limited area makes complex thermal hardware difficult to integrate.

\thermosaic\ works by dividing the fingertip contact area into multiple independently controlled thermal subregions and coordinating them over time. By shifting part of the skin toward an upcoming target temperature while preserving the current overall percept, the remaining physical change required for a perceptual category transition can be reduced. We draw on two established properties of thermal perception: spatial summation, which integrates distributed thermal inputs into a unified percept~\cite{HardyOppel1937_JCI_ThermalSummation,cataldo2016thermal}, and thermal adaptation, which reduces sensitivity to ongoing or gradual local changes~\cite{kenshalo1966temporal}. Together, these effects allow localized regions to be pre-warmed or pre-cooled without immediately disrupting the dominant sensation, enabling earlier perception of the final transition compared to a uniform actuator.

Our approach is conceptually related to prior work by Sato et al.~\cite{Sato_2012, Sato_2013jrm}, which provided an early indication that spatially heterogeneous stimulation can shape the temporal experience of temperature under relatively small skin-temperature changes (\SIrange{2}{4}{\celsius}).
\thermosaic\ extends this by systematically combining spatial summation and thermal adaptation to calibrated transitions across the full \SIrange{20}{40}{\celsius} operating range, reframing the problem as scheduled transitions among participant-specific cold, neutral, and hot perceptual categories.

To validate the perceptual basis of \thermosaic, we first developed a controlled benchtop 2$\times$2 fingertip thermal experimental setup with independently controlled miniaturized Peltier elements. Using this setup, we conducted three controlled perceptual studies to test whether: (1) spatially distributed thermal deviations are integrated into a single categorical percept, (2) thermal adaptation stabilizes that percept when part of the stimulus drifts toward neutrality, and (3) combining these effects reduces perceived transition time between thermal states. Across these studies, we found that spatially distributed stimulation can be integrated and adaptively maintained to reduce perceived transition time between thermal states. We then translated the same design principles into a standalone wearable implementation of \thermosaic\ and evaluated it in a VR scenario against a baseline thermal interface, showing that it significantly reduces perceived thermal lag and improves temporal alignment in interactive use.

In summary, our contributions are:
(1) We introduce \thermosaic, a spatiotemporal thermal feedback approach for accelerating perceived temperature transitions on the fingertip.
(2) Through three perceptual studies, we empirically show how spatial summation and thermal adaptation can be leveraged to reduce perceived transition time.
(3) We implement \thermosaic\ as a standalone wearable device and demonstrate in a VR study that it improves temporal alignment relative to a baseline thermal interface.

\section{Related Work}\label{sec:related_work}
\subsection{Thermal Interfaces in HCI}\label{subsec:thermal_interfaces_hci}

Thermal feedback has been widely explored in HCI as a way to simulate material properties~\cite{material_discrimination, yang2009spatial}, enhance immersion in virtual environments~\cite{fieryhands, let_it_snow, thermalgrasp_alex, DexteriSync}, and convey affective or communicative cues~\cite{multi_moji, thermal_taste, emotion_image, emotion_vib, thermon, thermo_photo}.

The most widely used actuator in thermal interfaces is the Peltier module, which generates heating or cooling by reversing the direction of the applied current. 
Its compact form factor and bidirectional heating and cooling capability make it suitable for a wide range of applications, including fingertip displays~\cite{fingertipthermal, nakatani2016novel, Sato2016} and smart rings~\cite{ZHU2019234} for the fingers; gloves~\cite{fieryhands, kim2020thermal, thermograsp}, wrist-worn~\cite{thermalbracelet, thermal_in_motion, ThermOuch, HeatFlow} and handheld devices~\cite{thermalcane} for the arms and hands; head- or face-mounted devices~\cite{face_on, thermoquest, ThermoVR}; torso-worn systems~\cite{ThermalWear}; and even ear- or nose-mounted interfaces~\cite{thermearhook, Trigeminal, augmentedbreathing}.

Beyond thermoelectric actuation, prior work has also explored pneumatic systems that circulate heated or cooled air~\cite{ThermAirGlove, chen2026soft}, liquid-based thermal garments~\cite{Therminator}, mid-air ultrasound heating~\cite{mid_air_thermo, non_contact_thermal}, hybrid hydro-optical techniques combining light and water~\cite{Hydroptical}, and chemical or physiological approaches such as alcohol evaporation cooling~\cite{ALCool}, trigeminal stimulation~\cite{Trigeminal}, and pain-based thermal cues~\cite{Douleur}.

Despite this diversity of actuation methods, the responsiveness of thermal interfaces remains fundamentally constrained by the slow dynamics of heat transfer. Conventional Peltier-based systems often require several seconds to produce perceptible temperature changes~\cite{fingertipthermal}, which can lead to temporal misalignment between thermal feedback and interaction events. To address this limitation, prior work has primarily focused on accelerating physical temperature change through hardware innovation. For example, ThermAirGlove~\cite{ThermAirGlove} increases cooling speed via forced air circulation, while Flip-Pelt~\cite{flip_pelt} enables rapid bidirectional transitions by mechanically alternating between preconditioned Peltiers. Although effective, these methods often rely on large hardware assemblies or moving parts, which can limit where thermal feedback can be deployed. This constraint is especially acute at the fingertip, where available space is highly limited.

Our work also uses Peltier modules as the primary thermal source, but approaches thermal latency differently. Rather than accelerating physical temperature change through additional hardware complexity, we investigate whether perceived thermal transitions can be accelerated through perceptual design.

\subsection{Psychophysics of Thermal Perception}\label{subsec:psychophysics_perception}
Thermal perception is shaped by both spatial and temporal mechanisms~\cite{filingeri2016neurophysiology, Sakurai2014, Sakurai2016}. 
A classic effect is thermal spatial summation~\cite{HardyOppel1937_JCI_ThermalSummation, summation1967, Marks1974, rozsa1977spatialsummation}, in which thermal inputs distributed across the skin are integrated into a unified percept. 
Increasing the stimulated area lowers detection thresholds and enhances perceived intensity~\cite{yang2009spatial}, while multi-point stimulation can bias the perceived temperature of a target location. 
When adjacent skin regions are exposed to different temperatures, perception integrates these inputs rather than represents them independently, often resulting in a single categorical sensation influenced by the overall spatial pattern~\cite{hot_cold_confusion}.
This integration process can reduce sensitivity to local discrepancies, allowing heterogeneous thermal inputs to be perceived as a coherent percept.

Another key mechanism is thermal adaptation~\cite{thermal_adaptation1, thermal_adaptation2, thermal_adaptation3, Akiyama2012}, which shifts perceptual sensitivity over time.
After sustained exposure to a given temperature, detection thresholds change, reducing sensitivity to subsequent stimuli in the same direction~\cite{kenshalo1968warm}. 
As a result, gradual temperature changes may remain perceptually unnoticed even when the underlying physical temperature has already shifted.
These adaptive shifts suggest that intermediate thermal states can sometimes be traversed without immediately altering the dominant percept.

Together, spatial summation and thermal adaptation suggest that thermal perception can be modulated through structured stimulation. 
These mechanisms create opportunities to bias overall thermal judgments and reduce the apparent perceptual distance between thermal states through spatiotemporal design.

A related direction was explored by Sato et al.~\cite{Sato_2012, Sato_2013jrm}, who showed that spatially distributed warm and cool stimulation enabled faster thermal perception than uniform stimulation when the skin-temperature change was approximately \SIrange{2}{4}{\celsius}. 
Their demonstration established the promise of spatial thermal design, but several questions remained. The pre-stimulus was separated from the main stimulus by a no-contact interval, making it unclear whether anticipation contributed to the effect. In addition, their evaluations were exploratory, and the apparatus was table-mounted, unsuited to interactive use.

\thermosaic\ extends this direction to participant-calibrated, category-preserving transitions across the \SIrange{20}{40}{\celsius} safe operating range. 
Continuous-contact preconditioning allows us to test conscious detectability directly. 
Through three powered perceptual studies, we systematically investigate how spatial summation and thermal adaptation can be combined. 
We further realize this approach in the first standalone fingertip-worn device, approximately 6.7 times smaller than Sato et al.'s table-mounted prototype, and demonstrate its use for accelerating perceived thermal transitions in an interactive VR scenario.

\section{Perceptual Basis and Study Logic}
\thermosaic\ is based on the idea that accelerating perceived thermal transitions does not require every part of the fingertip to change temperature at the same time. Instead, if some subregions can be shifted toward an upcoming target state while the overall percept remains stable, the remaining physical change required for the final perceptual transition may be reduced. In this way, the goal is not to accelerate physical temperature change itself, but to accelerate the moment at which a new thermal state is perceived.

In this work, fingertip thermal perception is treated in terms of three coarse perceptual categories: cold, neutral, and hot. These categories are not separated by sharp boundaries, but by fuzzy transition zones (Figure~\ref{fig:teaser}d)~\cite{schweiker2018drivers}. 
A thermal transition is therefore defined here not as a precise change in physical temperature, but as the moment when the overall percept crosses from one category into another. \thermosaic\ operates by shifting subregions of the fingertip toward an upcoming target category while preserving the dominant overall percept in the current one.

As discussed in Section~\ref{subsec:psychophysics_perception}, this strategy depends on two perceptual conditions. First, spatial summation must be sufficient to integrate heterogeneous thermal inputs across the fingertip into a single dominant percept rather than exposing each local deviation independently. If this holds, local subregions can deviate from one another while the overall sensation remains within the same perceptual category. Second, thermal adaptation must be sufficient to allow partial local temperature changes to occur without immediately altering that dominant percept. If this holds, some subregions can be gradually shifted toward a target category while the user still perceives the original state. When these two conditions are combined, subregions of the fingertip may be shifted toward an upcoming target state before the user becomes aware of a transition. The remaining temperature change required to cross the perceptual boundary is then reduced, which may shorten the perceived transition time between categories.

Three perceptual studies were conducted to test these perceptual assumptions. Study~1 examines whether spatially distributed thermal deviations on the fingertip are integrated into a stable overall percept of hot, neutral, or cold. Study~2 examines whether, after sustained exposure, part of the stimulus can drift toward neutrality while the dominant percept remains unchanged. Study~3 then investigates whether combining these effects shortens perceived transition time across thermal categories.
To support these perceptual studies, we developed a benchtop 2$\times$2 fingertip thermal setup with independently controlled miniaturized Peltier elements. We use this setup to systematically manipulate thermal subregions and test the perceptual basis under controlled conditions, before translating the same design logic into a standalone wearable implementation.

\section{Benchtop Experimental Setup}\label{sec: benchtop}
\begin{figure}[h]
  \centering
  \includegraphics[width=1\linewidth]{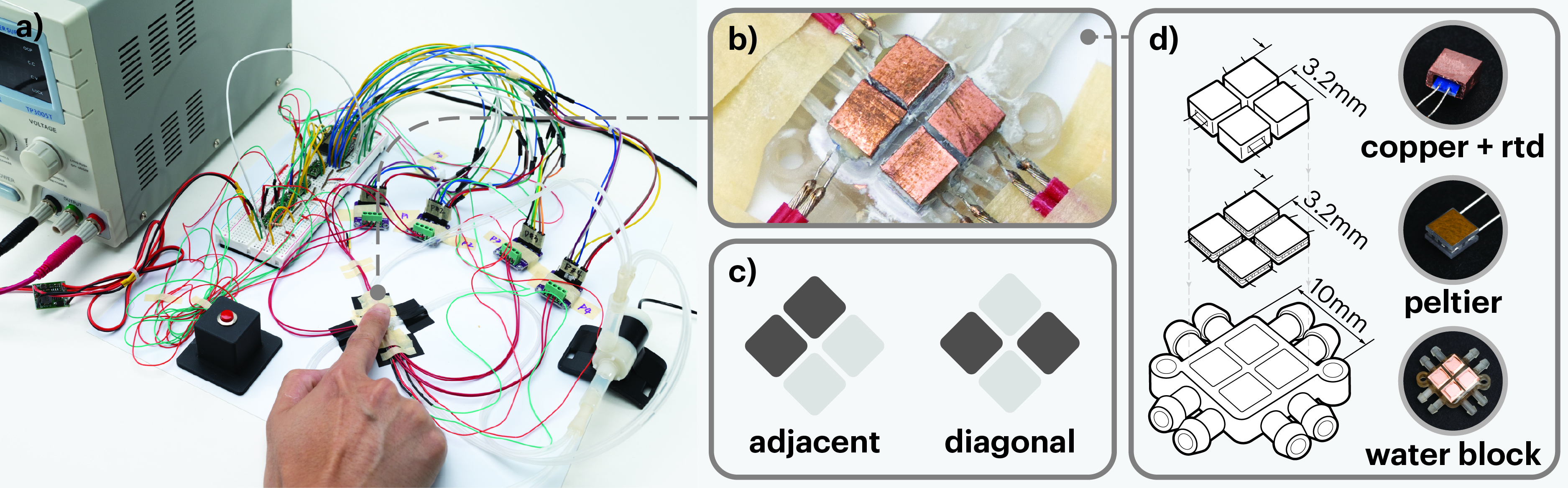}
  \caption{(a) Benchtop setup for fingertip thermal stimulation.
  (b) Close-up of the fingertip thermal actuator assembly.
  (c) Thermal activation patterns.
  (d) Exploded schematic of the fingertip thermal actuator, comprising four resistance temperature detectors (RTDs) embedded in copper blocks, four Peltier modules, and a water block.}
  \Description{A multi-panel figure illustrating a fingertip thermal stimulation system. Panel (a) shows a benchtop experimental setup with a power supply, control electronics, wiring, and a fingertip device being touched by a user. Panel (b) presents a close-up view of the fingertip thermal actuator assembly, highlighting four small copper blocks arranged in a square with electrical connections. Panel (c) depicts two example thermal activation patterns labeled adjacent and diagonal, indicating different combinations of activated elements. Panel (d) shows an exploded schematic of the actuator, including four resistance temperature detectors embedded in copper blocks, four Peltier modules beneath them, and a water block used for heat dissipation.}
  \label{fig:benchtop_setup}
\end{figure}

To test the perceptual basis of \thermosaic, we developed a benchtop 2$\times$2 fingertip thermal setup (Figure~\ref{fig:benchtop_setup}a) composed of four independently controlled miniaturized Peltier elements. The setup enables systematic manipulation of local thermal subregions on the fingertip while maintaining sufficient spatial resolution for the perceptual studies described in Section~3.

\subsection{Design Rationale}
The setup partitions the fingertip contact area into four independently controlled thermal subregions. A 2$\times$2 layout was chosen as the smallest configuration that enables spatially non-uniform thermal patterns while remaining compact enough for stable fingertip contact. This partition provides sufficient degrees of freedom to test the two perceptual conditions required by \thermosaic: whether heterogeneous local inputs are integrated into a single dominant percept, and whether subregions of the contact area can be shifted without immediately altering that percept.

To determine the dimensions of the array, we approximated the contactable fingertip pad as an ellipse with a width of \SI{10}{\milli\meter} and a height of \SI{13}{\milli\meter}. The active contact area was then modeled as a $45^{\circ}$-rotated square inscribed within this ellipse, yielding an effective side length of approximately \SI{7.07}{\milli\meter}. Dividing this area into a 2$\times$2 layout gives subregions of roughly \SI{3.5}{\milli\meter} per side.
We therefore selected the 3.2~mm \(\times\) 3.2~mm Peltier modules to closely match the target subregion dimensions.

The 2$\times$2 layout offers two possible pairwise configurations for generating non-uniform thermal patterns (Figure~\ref{fig:benchtop_setup}c): adjacent and diagonal. 
Through formative observations, we found that both pairings produced broadly similar percepts. We therefore adopted diagonal grouping, which provides a more balanced spatial distribution and enables two interleaved thermal groups to be driven toward different temperatures without forming a localized cluster.

\subsection{Hardware Implementation}
The experimental setup consisted of four Peltier modules (Custom Thermoelectric LLC, model 00701-9A30-14RU3) arranged in a 2$\times$2 diamond array (Figure~\ref{fig:benchtop_setup}b). 
Each Peltier was topped with a CNC-milled copper block of matching size, which served as the thermal contact layer and contained a pocket for an embedded resistance temperature detector (RTD; Yageo Nexensos, model 5157701) to measure the fingertip--Peltier interface temperature without obstructing skin contact. 
A customized 3D-printed water block beneath the array independently cooled each Peltier in a closed, reservoir-free loop containing approximately \(12~\mathrm{mL}\) of working fluid, reducing thermal interference (Figure~\ref{fig:benchtop_setup}d).

Thermally conductive adhesive (MG Chemicals, model 8329TFF) was applied at the interfaces between the copper--RTD, copper--Peltier, and Peltier--water block to ensure mechanical bonding and enhance heat transfer. The assembled device was secured to a flat surface during the studies to ensure stable operation and consistent fingertip--device contact.

\subsection{Thermal Control and Performance}
\label{sec:thermal_performance}

Each Peltier module was regulated independently by a PID controller that continuously adjusted PWM duty cycle to minimize the error between target and measured temperature. Because heating and cooling exhibit different thermal dynamics, we used asymmetric PID control with separate gain parameters for heating and cooling. Tuning was initialized using the Ziegler--Nichols method and refined manually to balance responsiveness and stability.

To ensure safe and consistent operation, actuator outputs were constrained to a \SIrange{20}{40}{\celsius} range, spanning the perceptually relevant thermal categories while preventing discomfort~\cite{10.1093/annhyg/mel030, MARTIN20171624}. An overheating protection mechanism also monitored each module's surface temperature in real time and automatically reduced or suspended PWM output if excessive heating was detected.

The thermal dynamics of the setup were characterized using repeated step responses. The mean heating and cooling rates were \SI{2.04}{\celsius\per\second} ($\mathrm{SD}=0.24$) and \SI{2.07}{\celsius\per\second} ($\mathrm{SD}=0.22$), respectively. Steady-state error remained within \SI{\pm 0.3}{\celsius} across modules, with a maximum deviation below \SI{0.7}{\celsius}. Additional ramp-rate data are provided in Appendix~\ref{app:ramp_data}.

\section{Thermal Perceptual Studies}
\label{sec:perception_study}
To test the perceptual basis of \thermosaic, we conducted three perceptual studies within a single experimental session. 
The studies progress from spatial integration, to maintenance of a dominant percept under partial thermal change, and finally to perceived transition time between thermal categories.
All studies were approved by our institution’s Institutional Review Board (IRB).

\subsection{Participants}
We recruited 15 participants (7 male, 8 female) with a mean age of 26.7 years (SD = 2.87). All reported normal thermal sensitivity and no injuries to their fingertips or hands. Fourteen were right-hand dominant. All participants provided written informed consent and received US \$40 as compensation.

\subsubsection{Power Analysis and Sample Size Justification.}
All three studies were conducted with the same group of participants. 
Thus, the sample size was determined based on the most demanding statistical requirements across studies. Study~1 involved near-ceiling categorical identification, where power analysis is not informative. The design therefore emphasized repeated observations per participant.
For Study~2, power was estimated via simulation using a generalized linear mixed-effects model matching the planned analysis. A conservative detection probability ($p_{\text{detect}} = 0.05$) was assumed to reflect near-chance detection under masking.
For Study~3, simulations matched the planned linear mixed-effects model and targeted a minimal meaningful effect corresponding to a 1.0\,s reduction in detection time.
Both analyses indicated that $N=12$ participants achieved power greater than 0.80.
We therefore recruited $N=15$ participants to provide a margin above the minimum required sample size, allowing for potential exclusions and supporting stable mixed-effects estimation across studies.

\subsection{Shared Experimental Protocol}\label{subsec: shared protocol}
All perceptual studies were conducted with the same participant group using a within-subjects design and the benchtop experimental setup in Section~\ref{sec: benchtop}. The studies took place in a quiet laboratory room maintained at \SI{22 \pm 1}{\celsius}. 
Participants were seated at a desk with both hands placed on an armrest to stabilize fingertip contact. No external airflow or heat sources were present. 

Before Study 1, we calibrated four thermal category thresholds for each participant, 
$C$ (cold anchor), $N_1$ (cold–neutral boundary), $N_2$ (neutral–hot boundary), and $H$ (hot anchor)
(Figure~\ref{fig:teaser}d, see Appendix~\ref{app:thermal_cali} for details of the calibration procedure). 
Three temperature categories, \textit{cold}, \textit{neutral}, and \textit{hot}, were then defined based on these participant-specific threshold boundaries.
This calibration accounted for inter-individual variability in thermal sensitivity and established more comparable perceptual conditions across participants than applying identical absolute temperatures~\cite{ADAMCZYK20221823, DEOLINDO2025104732}. Similar procedures have been widely used to match subjective sensation levels in psychophysical and applied thermal studies~\cite{weik2022placebo, roquet2021interoceptive, duan2025real}. The resulting calibrated values were used throughout all three studies.

In all studies, participants used the index fingers of both hands. The index fingertip was chosen as the test site because it is among the most thermally sensitive regions of the hand~\cite{joseph_sensitivity}, and sensitivity differences between the left and right fingertips are negligible~\cite{moloney2011reliability}. Participants were instructed to alternate between the left and right index fingertips across trials to provide additional recovery time for each fingertip and help reduce local fatigue. The starting hand was counterbalanced across participants.

\subsection{Study 1: Spatial Integration of Heterogeneous Thermal Inputs}

Study~1 tested whether spatially heterogeneous stimulation across the diagonally split pattern is integrated into a single categorical percept on the fingertip. 
We hypothesized that, compared to the baseline condition in which all four Peltier modules produced a uniform temperature, the diagonally split pattern would preserve categorical identification of cold, neutral, and hot sensations despite local temperature differences across subregions.

\subsubsection{Study Conditions}
After the calibration procedure described in Section~\ref{subsec: shared protocol}, the four participant-specific thermal category thresholds ($C$, $N_1$, $N_2$, $H$) were used to define the study conditions.

For each thermal category, we compared a uniform pattern (baseline) with a diagonally split pattern (test condition), as shown in Figure~\ref{fig:study1_result}a.
In the baseline condition, all four Peltier modules were set to the same temperature: \textit{cold} was set to $C$, \textit{hot} was set to $H$, and \textit{neutral} was defined as the mean of $N_1$ and $N_2$.

The test conditions implemented the proposed diagonally split pattern by assigning different temperatures to the two diagonal groups. All temperatures are reported in \si{\celsius}. For \textit{neutral}, we used two heterogeneous variants: one with the diagonals set to $N_1+1$ and $N_2+2$, and one with the diagonals set to $N_1-2$ and $N_2-1$. For \textit{cold}, the two diagonals were set to 20 and $C+2$. For \textit{hot}, the two diagonals were set to 40 and $H-2$. 
These values were chosen to create pronounced local temperature deviations while keeping the overall pattern near the intended thermal category. 
Milder alternatives would therefore be expected to cause less disruption.
Specifically, 20 and 40 represented the lower and upper actuator bounds used in the setup, while the offsets around $N_1$, $N_2$, $C$, and $H$ introduced modest within-category deviations.

\subsubsection{Procedure}

Before the main experiment, participants completed two sample trials to familiarize themselves with the sensations and the response procedure.

To minimize temporal adaptation and fatigue, a permuted-block randomization was used: the seven experimental conditions were grouped into a super-block, randomly ordered within each block, and repeated three times, yielding 21 trials per participant. 
For each test condition involving two temperatures, the assignment of temperatures to the two diagonals of the 2$\times$2 array was counterbalanced across participants.
Hand order followed the shared experimental protocol described in Section~\ref{subsec: shared protocol}.

In each trial, participants placed the designated fingertip in full contact with the four copper surfaces for 4 seconds. 
After lifting their finger, they reported the perceived temperature by selecting the category that best matched their sensation and rating it with the highest confidence. 
To reduce thermal adaptation, a 10-second rest interval was inserted between trials, and a two-minute break was provided between super-blocks. 
Including introduction, calibration, and debriefing, the study lasted approximately 20–25 minutes.

\subsubsection{Results and Analysis}
The primary outcome measure is categorical identification accuracy. 
Across all trials, participants reported experiencing a single, uniform temperature sensation. None was able to detect or distinguish any localized temperature variation among the four modules.

\begin{figure}[h]
  \centering
  \includegraphics[width=\linewidth]{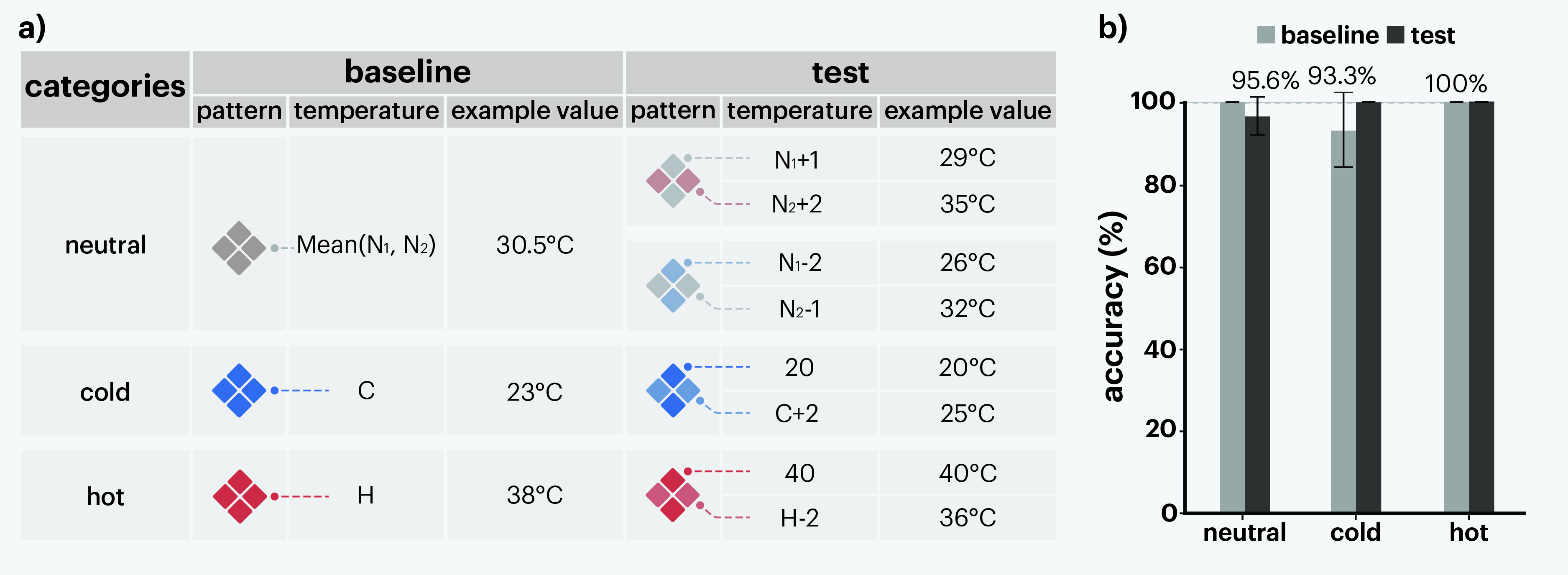}
  \caption{Study 1 experimental conditions and outcomes. (a) Applied temperature profiles for each condition. (b) Accuracy of categorizing perceived thermal sensations.}
  \Description{Panel (a) illustrates the applied temperature profiles across experimental conditions. Panel (b) presents the accuracy with which participants categorized their perceived thermal sensations under the baseline and test conditions.}
  \label{fig:study1_result}
\end{figure}

Results are shown in Figure~\ref{fig:study1_result}b. For \textit{neutral} stimuli, accuracy reached 100\% (SD = 0\%) in the baseline condition and remained comparably high in the test condition at 95.6\% (SD = 7.6\%). For \textit{cold} stimuli, mean accuracy was 93.3\% (SD = 13.8\%) in the baseline condition and 100\% (SD = 0\%) in the test condition. For \textit{hot} stimuli, accuracy was consistently 100\% (SD = 0\%) in both the baseline and the test conditions.

Error patterns were further examined using confusion matrices. In the baseline cold condition, the majority of errors occurred when cold was misclassified as neutral. In contrast, in the test cold condition, no such errors were observed. Errors in neutral and hot conditions were rare and not systematically biased toward any single category.

As participants performed the categorization task with near-perfect accuracy across all conditions, we also computed accuracy for each participant as the proportion of correctly identified trials in the baseline and test conditions. To evaluate the effect of the diagonal actuation pattern, we analyzed the paired accuracy difference between the test and the baseline condition.
\begin{equation}
\Delta p = p_{\text{test}} - p_{\text{baseline}}
\end{equation}

Confidence intervals for $\Delta p$ were estimated using bootstrap resampling across participants with 20{,}000 iterations, providing distribution-free uncertainty estimates appropriate for near-ceiling performance. 
The mean paired accuracy difference between the test and the baseline was $0.0$, with a $95\%$ bootstrap confidence interval of $[-3.1\%,\,3.3\%]$. This interval lies entirely within the predefined equivalence margin ($\Delta = 5\%$), supporting both equivalence and non-inferiority of the diagonally split pattern (test condition) relative to the uniform pattern (baseline).

These results indicate that the diagonally split pattern preserves categorical temperature identification, consistent with stable thermal perceptual categories despite spatially deviant actuation.

\subsection{Study 2: Maintenance of a Dominant Thermal Percept}

Study~2 tested whether a dominant hot or cold percept could be maintained when part of the fingertip input was shifted toward neutrality. We hypothesized that, after sustained exposure to an initial thermal category, partial local changes would not necessarily be reported as a perceptual transition, especially when the thermal pattern remained spatially balanced across the fingertip.

\subsubsection{Study Conditions}

Each trial consisted of two phases, as shown in Figure~\ref{fig:study2_result}a.
In Phase~1 (0--10~s), the Peltier array rendered an initial thermal pattern to establish a dominant percept. In Phase~2 (10--20~s), a second thermal pattern was presented to test whether participants would report a perceptual change between the two phases.

Several conditions were defined. Condition~1 served as a no-change baseline, where the full array remained neutral across the two phases. Conditions~2 and 3 served as large-change baselines: the array began in a cold or hot pattern in Phase~1 and changed uniformly to neutral in Phase~2.
These conditions were expected to elicit reports of temperature change and therefore served as sensitivity controls.

Conditions~4--7 were the test conditions. In Conditions~4 and~5, Phase~1 established a cold percept. In Phase~2, the pattern was partially shifted toward the upper boundary of the neutral zone, closest to hot, to test whether this partial shift would trigger a noticeable perceptual change. Condition~4 changed only the diagonal group at \(C+2\), raising it to \(N_2\), while the diagonal group at 20 remained unchanged. Condition~5 changed both diagonal groups, lowering the group at \(C+2\) to 20 and raising the group at 20 to \(N_2\). Conditions~6 and~7 followed the same logic from a hot initial state. In Phase~2, the pattern was shifted toward the lower boundary of the neutral zone, closest to cold, to test whether a noticeable perceptual change would be detected.

Both cooling and heating ramp rates were fixed at \SI{2}{\celsius\per\second} across all conditions. 
Each trial lasted 20 seconds. 
The initial pattern was maintained during Phase 1 (0--10~s), after which the Phase~2 pattern was applied and maintained for the remaining 10 seconds based on the assigned condition.

\subsubsection{Procedure}
 
Following the same block structure as in Study 1, the seven experimental conditions were grouped into a super-block, randomized within each block, and repeated three times, resulting in 21 trials per participant. 
The temperature assignments for the two diagonal groups were counterbalanced across participants. Starting hand was counterbalanced between participants as described in Section~\ref{subsec: shared protocol}. 
Before the main experiment, participants completed two practice trials to familiarize themselves with the task.

In each trial, participants placed the designated fingertip in full contact with the four copper surfaces. 
They were informed of the initial temperature category and then maintained contact for 20 seconds. 
After lifting their finger, they reported whether they had noticed any temperature change during the trial and, if so, identified the final temperature category they perceived. 
A 20-second rest interval was inserted between trials, and a three-minute break was provided between super-blocks. 
The study lasted approximately 40-45 minutes.

\subsubsection{Results and Analysis}

\begin{figure}[h]
  \centering
  \includegraphics[width=\linewidth]{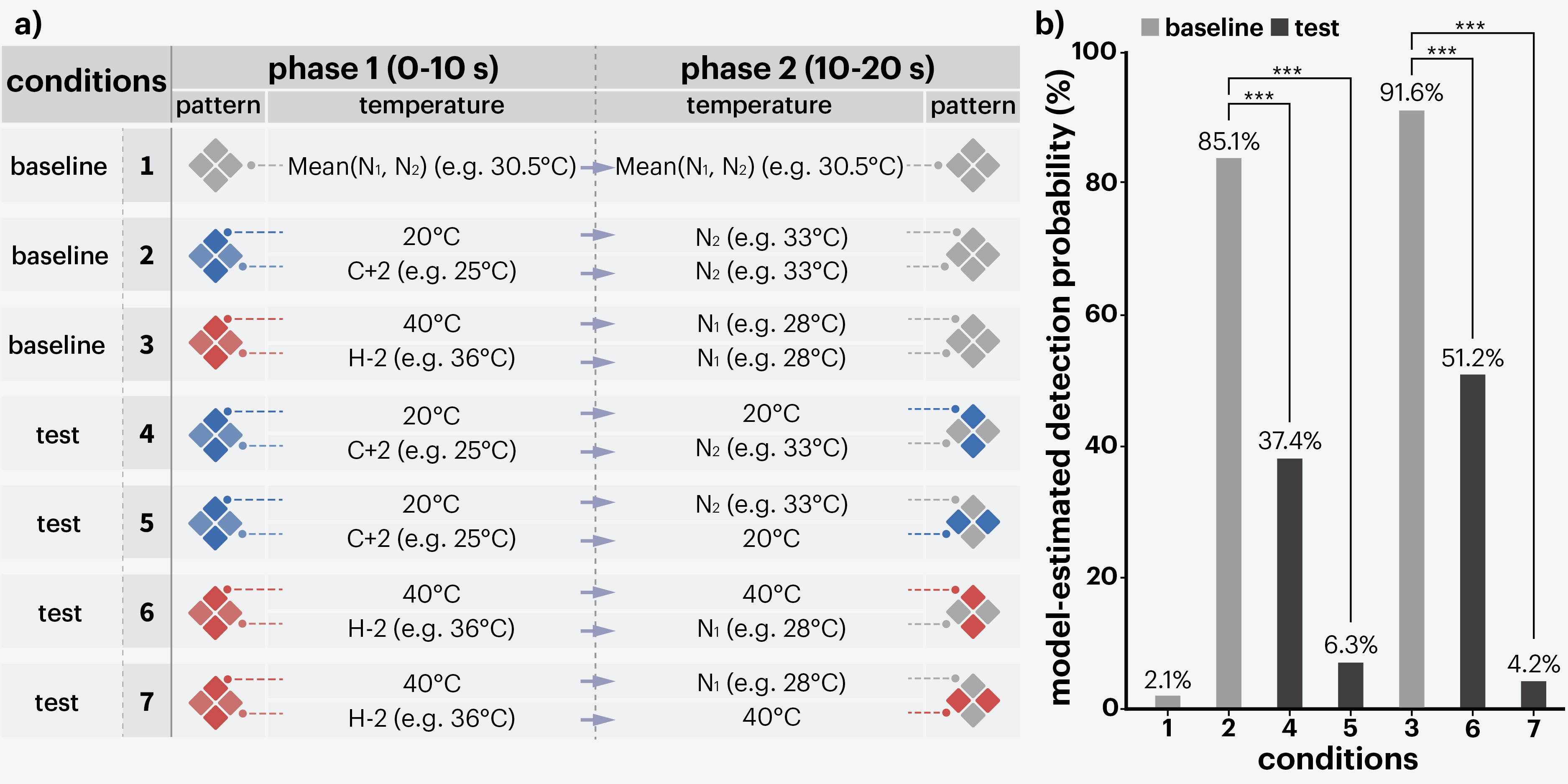}
  \caption{Study 2 experimental conditions and outcomes. (a) Applied temperature profiles for each condition. (b) Model-estimated probabilities of reporting a temperature change across conditions.}
  \Description{Panel (a) shows applied temperature profiles for each condition. Panel (b) bar chart showing model-estimated probabilities of reporting a temperature change across conditions. Brackets indicate model-based contrasts with the corresponding large-change control.}
  \label{fig:study2_result}
\end{figure}

Detection responses were analyzed using a logistic mixed-effects model with Condition as a fixed effect and participant as a random intercept. 
We report model-estimated detection probabilities with 95\% confidence intervals, and model-based odds ratios (OR) as effect-size estimates.
Results are shown in Figure~\ref{fig:study2_result}b.

In the no-change baseline (Condition~1), the estimated probability of reporting a temperature change was low (2.1\%, 95\% CI [0.3\%, 13.6\%]), indicating a low false-alarm rate. 
In contrast, the large-change controls produced high detection probabilities: 85.1\% [71.0\%, 93.0\%] for cold-to-neutral (Condition~2) and 91.6\% [79.9\%, 96.9\%] for hot-to-neutral (Condition~3). 
These results confirm that participants reliably detected salient temperature transitions.

The balanced-shift conditions produced substantially lower detection probabilities. 
For cold-origin trials, the estimated probability was 6.3\% [2.0\%, 18.3\%] (Condition~5), compared with 85.1\% in the corresponding large-change control, a large reduction in the odds of reporting a change (OR = 0.012, 95\% CI [0.003, 0.051], $p < .001$). 
For hot-origin trials, the estimated probability was 4.2\% [1.0\%, 15.6\%] (Condition~7) compared with 91.6\% in the corresponding control (OR = 0.004, 95\% CI [0.001, 0.024], $p < .001$). 
Under balanced shifts, participants thus largely maintained the original dominant percept despite substantial changes in the underlying stimulation.

The unilateral-shift conditions showed intermediate detection probabilities: 37.4\% [23.8\%, 53.2\%] for cold-origin trials (Condition~4) and 51.2\% [35.9\%, 66.2\%] for hot-origin trials (Condition~6). 
Although these conditions also reduced the odds of reporting a change relative to the large-change controls (cold-origin: OR = 0.104, 95\% CI [0.037, 0.294], $p < .001$; hot-origin: OR = 0.096, 95\% CI [0.029, 0.320], $p < .001$), their detection probabilities remained much higher than those of the balanced-shift conditions.

As a complementary descriptive analysis, we computed signal-detection sensitivity ($d'$) using the no-change condition (Condition~1) as the shared false-alarm baseline, with log-linear correction. 
The large-change controls showed high sensitivity (cold-origin: $d' = 2.83$; hot-origin: $d' = 3.14$). 
Balanced-shift conditions showed low sensitivity (cold-origin: $d' = 0.41$; hot-origin: $d' = 0.24$), consistent with these changes being largely masked by the dominant thermal percept, while unilateral shifts produced intermediate values (cold-origin: $d' = 1.54$; hot-origin: $d' = 1.87$).

Taken together, these findings suggest that perceptual stability is more reliably preserved when changes in thermal input are spatially balanced. Specifically, when temperatures in the diagonal regions shifted in opposing directions, spatial integration maintained a dominant categorical percept, and gradual local changes largely went unnoticed. This pattern is consistent with an adaptation-based account in which sensitivity to covert changes is reduced when the overall spatial structure of the thermal pattern remains stable.

\subsection{Study 3: Perceived Transition Time Between Thermal Categories}

Study~3 tested the core claim of \thermosaic: whether combining spatial integration and adaptation can shorten the time required to perceive a transition between thermal categories. Building on Study~2, we hypothesized that the effect would be strongest when the initial state was already categorically established as cold or hot, and weaker or absent when the initial state was neutral.

\subsubsection{Study Conditions}
We examined all ordered transitions between the three temperature categories, yielding six transition conditions: neutral-to-hot, neutral-to-cold, cold-to-neutral, cold-to-hot, hot-to-neutral, and hot-to-cold (Figure~\ref{fig:study3_result}a). The six transition conditions were instantiated using the participant-specific thresholds established in Section~\ref{subsec: shared protocol}.

Each trial consisted of three phases. In Phase~1 (0--5~s), the array rendered a stable thermal pattern corresponding to the starting category. In Phase~2 (5--10~s), the baseline condition maintained this initial pattern, whereas the test condition applied the proposed diagonal preconditioning pattern. In Phase~3 (10--20~s), both baseline and test conditions transitioned to the target category. 

In the baseline condition, all four Peltier modules were driven uniformly. Building on the manipulations introduced in Studies~1 and~2, the test condition applied a diagonally split pattern that introduced a partial, spatially heterogeneous shift during Phase~2 before the full transition occurred in Phase~3.

\subsubsection{Procedure}
The study was divided into two parts. In Part~1, all trials began at a neutral temperature. In Part 2, trials began at either a cold or a hot temperature.
Participants completed two practice trials to familiarize themselves with the procedure before the main experiment.
The starting hand was counterbalanced between participants as described in Section~\ref{subsec: shared protocol}.

Before each trial, participants were informed of the transition type. 
They then placed the index finger on the device. Once they confirmed that the rendered sensation matched the instructed initial state, the 20-second trial began.
Participants were instructed to press a button as soon as they detected a shift to the target temperature.
Response time was recorded automatically as the interval between the programmed onset of the temperature shift and the button press.

Following the same block structure as in Study 1,
Part 1 grouped four conditions into a super-block, randomized within each block and repeated three times, yielding 12 trials per participant. 
Part 2 included eight experimental conditions per super-block, randomized within each block, and repeated three times, resulting in 24 trials per participant. 
A 30-second rest interval was provided between trials, a three-minute break between super-blocks, and a five-minute break between the two parts of the study.
In total, each participant completed 36 trials, and the study lasted approximately 50 minutes.

\subsubsection{Results and Analysis}
\begin{figure}[ht]
  \centering
  \includegraphics[width=\linewidth]{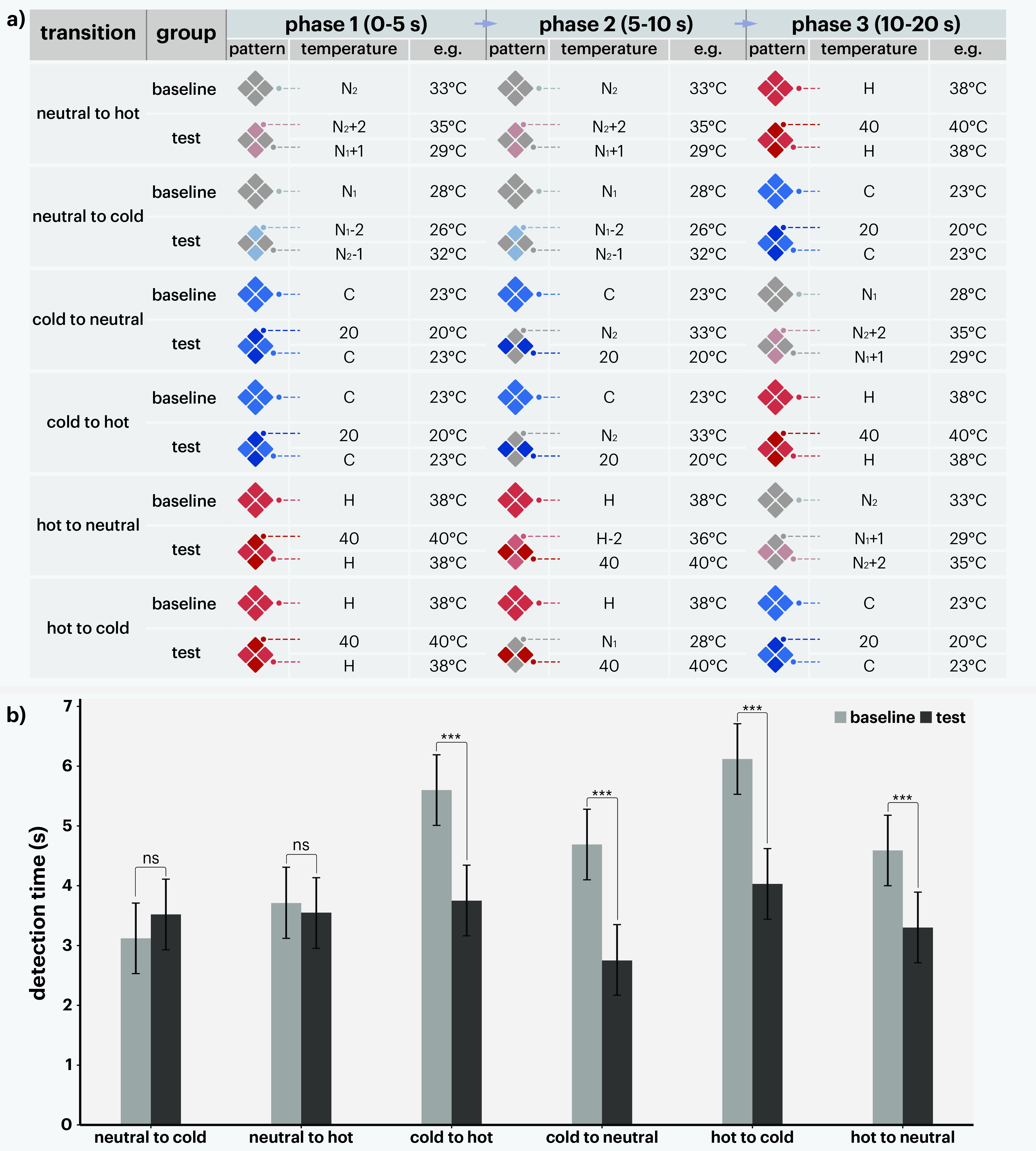}
  \caption{Study 3 experimental conditions and outcomes. (a) Applied temperature profiles for each condition. (b) Time to detect temperature transitions across different conditions.}
  \Description{Panel (a) shows applied temperature profiles for each condition. Panel (b) bar chart comparing the time required to detect temperature transitions across different conditions for baseline and test groups.}

  \label{fig:study3_result}
\end{figure}

Detection times between baseline and test conditions were analyzed using a linear mixed-effect model. 
The model included Group (baseline vs. test), Transition (six temperature transitions as shown in Figure~\ref{fig:study3_result}a), and their interaction as fixed effects, with participant treated as a random effect to account for repeated measures and individual differences. Degrees of freedom were estimated using the Kenward–Roger approximation.
As effect-size estimates for the omnibus effects, we report approximate partial eta-squared values ($\eta_p^2$) computed from the Type~III $F$ statistics.
Results are shown in Figure~\ref{fig:study3_result}b. 

This linear mixed-effects analysis revealed a significant main effect of Group
($F(1,514)=61.33$, $p<.001$, $\eta_p^2 = .107$), a significant main effect of Transition
($F(5,514)=13.91$, $p<.001$, $\eta_p^2 = .119$), and a significant Group~$\times$~Transition interaction
($F(5,514)=8.21$, $p<.001$, $\eta_p^2 = .074$), indicating that the effect of Group differed across Transition types.

Post-hoc comparisons based on estimated marginal means (Kenward–Roger correction) showed that detection times were consistently shorter in the test condition than in the baseline condition for transitions originating from non-neutral thermal states.

For cold-origin transitions, detection time decreased from 5.60~s in the baseline condition to 3.75~s in the test condition for cold-to-hot transitions (95\% CIs [5.01, 6.19] and [3.16, 4.34], respectively; $\Delta = 1.85$~s, $t = 5.11$, $p < .001$), and from 4.69~s in the baseline condition to 2.75~s in the test condition for cold-to-neutral transitions (95\% CIs [4.10, 5.28] and [2.16, 3.34], respectively; $\Delta = 1.94$~s, $t = 5.37$, $p < .001$).

Similarly, for hot-origin transitions, detection time decreased from 6.12~s in the baseline condition to 4.03~s in the test condition for hot-to-cold transitions (95\% CIs [5.53, 6.71] and [3.44, 4.62], respectively; $\Delta = 2.09$~s, $t = 5.77$, $p < .001$), and from 4.59~s in the baseline condition to 3.30~s in the test condition for hot-to-neutral transitions (95\% CIs [4.00, 5.18] and [2.71, 3.89], respectively; $\Delta = 1.29$~s, $t = 3.57$, $p < .001$).

In contrast, transitions originating from the neutral state showed no reliable differences between conditions. 
Detection times did not differ significantly between the baseline and test conditions for neutral-to-cold transitions (3.12~s vs. 3.52~s; 95\% CIs [2.53, 3.71] and [2.93, 4.11], respectively; $\Delta = -0.40$~s, $t = -1.10$, $p = .27$) or neutral-to-hot transitions (3.71~s vs. 3.55~s; 95\% CIs [3.12, 4.31] and [2.95, 4.14], respectively; $\Delta = 0.17$~s, $t = 0.47$, $p = .64$).

Overall, these results indicate that perceived transition time can be reduced when the initial thermal state is categorically established as cold or hot. In these cases, the test condition produced reliable reductions of approximately 1.29--2.09 seconds relative to baseline, corresponding to approximately 28\%--41\% faster detection. No comparable benefit was observed for transitions originating from the neutral state. This pattern is consistent with the proposed mechanism: preconditioning is most effective when it operates on top of an already dominant thermal percept.

\subsection{Summary}

Across the three studies, the results support the perceptual basis of \thermosaic. Study~1 shows that spatially heterogeneous fingertip stimulation can be integrated into a single categorical percept. Study~2 shows that a dominant hot or cold percept can often be maintained even when part of the stimulation shifts toward neutrality. Study~3 shows that combining these effects can shorten perceived transition time, particularly for transitions originating from an already established hot or cold percept.

\section{Wearable Implementation of \thermosaic}
\begin{figure}[h]
    \centering
    \includegraphics[width=\columnwidth]{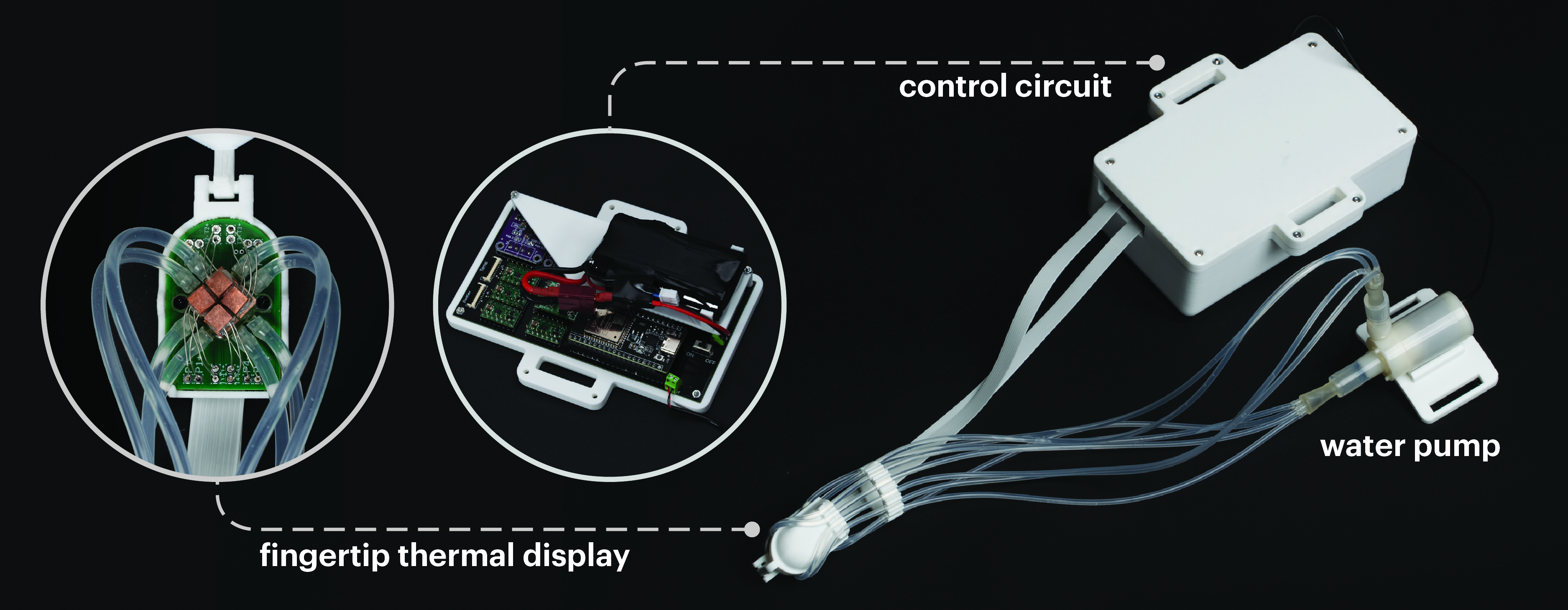}
    \caption{\thermosaic wearable implementation.}
    \Description{TherMosaic wearable implementation.}
    \label{fig:wearable_implementation}
\end{figure}

After validating the perceptual basis of \thermosaic\ with the benchtop experimental setup, we translated the same thermal core into a standalone fingertip-worn device (Figure~\ref{fig:wearable_implementation}) for interactive use. The wearable preserves the validated 2$\times$2 architecture used in the perceptual studies, including the same 3.2~mm \(\times\) 3.2~mm Peltier modules, copper blocks with embedded RTD sensors, and water-cooling blocks. This maintains a fingertip contact area of approximately 7~mm \(\times\) 7~mm while preserving the same independently controlled thermal subregions used in the benchtop studies.

To support wearable operation, we redesigned the surrounding hardware for standalone use. 
This included a fingertip-sized custom PCB for routing sensing and actuation signals, onboard control electronics, a rechargeable battery power supply, and a custom 3D-printed enclosure. 
The wearable uses the same water-cooling system and matches the benchtop’s ramp performance.
The resulting device preserves the validated thermal interface while enabling standalone fingertip wear for the interactive evaluation in Section~7.

\section{VR Evaluation}
To evaluate whether the wearable implementation of \thermosaic\ carries the perceptual advantage observed in Section~5 into interactive use, we conducted a VR study comparing \thermosaic\ against a uniform baseline condition. Participants played a mini-game involving repeated hot-to-cold and cold-to-hot transitions triggered by visual events. Our primary question was whether \thermosaic\ would improve the perceived temporal alignment between thermal and visual changes in VR. The study was approved by our institution's IRB.

\subsection{Participants}
Twelve participants (7 male, 5 female; mean age = 25.3, SD = 3.2) participated in the study. 
Ten were right-hand dominant. 
All reported normal thermal sensitivity and no hand injuries. 
Participants provided written informed consent and received US \$20 compensation.

\subsection{Apparatus and Procedure}
\begin{figure}[h]
    \centering
    \includegraphics[width=\columnwidth]{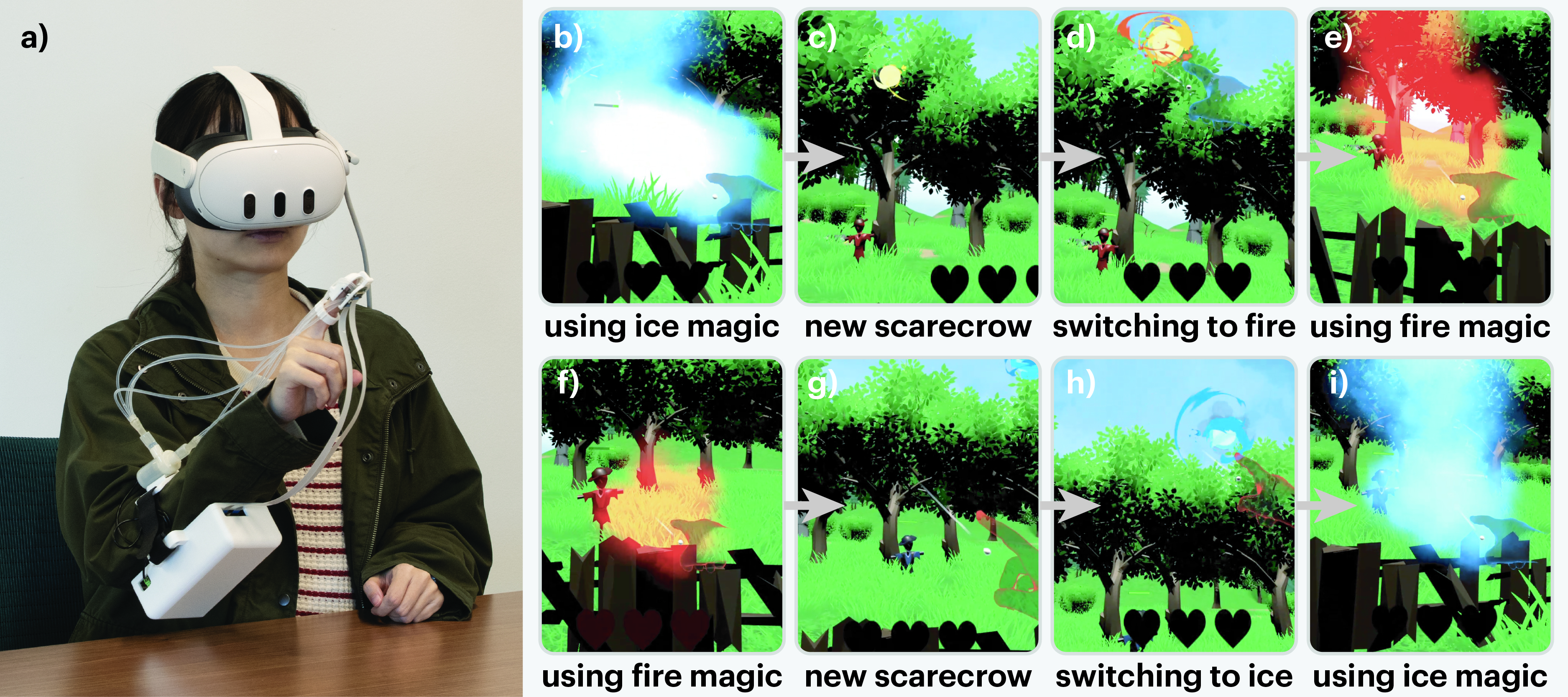}
    \caption{(a) Participant wearing the VR headset and \thermosaic device.
    (b-e) VR sequence after defeating a blue scarecrow: a new red scarecrow appears; the participant switches from ice magic to fire magic and uses fire magic to defeat the red scarecrow.
    (f-i) VR sequence after defeating a red scarecrow: a new blue scarecrow appears; the participant switches from fire magic to ice magic and uses ice magic to defeat the blue scarecrow.}
    \Description{Panel (a) A participant wearing a VR headset and TherMosaic device interacts with the system. Panel (b–e) After defeating a blue scarecrow, a red scarecrow appears; the participant switches from ice to fire magic and defeats it. Panel (f–i) After defeating a red scarecrow, a blue scarecrow appears; the participant switches from fire to ice magic and defeats it.}
    \label{fig:vrstudy}
\end{figure}

Participants wore a Meta Quest 3 VR headset and the \thermosaic wearable (Figure~\ref{fig:vrstudy}a) on the index finger. The water pump and control circuit were attached to the forearm using Velcro straps~\cite{jetunit}.

Prior to the study, we calibrated each participant's thermal thresholds to ensure perceptual consistency.
Across participants, the resulting thresholds (mean $\pm$ SD) were
\SI{22.67 \pm 0.49}{\celsius} for cold anchor,
\SI{27.83 \pm 0.94}{\celsius} for cold-neutral threshold,
\SI{33.00 \pm 0.74}{\celsius} for neutral-hot threshold,
and \SI{38.42 \pm 0.79}{\celsius} for hot anchor.
The device operated within a safe range of \SIrange{20}{40}{\celsius} with overheat protection. All thermal stimuli were generated with consistent ramp rates of \SI{2}{\celsius\per\second} across conditions.

After calibration, participants completed a brief training session to familiarize themselves with the interaction in VR. They then completed two gameplay rounds, one in the baseline condition and one in the \thermosaic\ condition, with order counterbalanced across participants. After each round, participants completed a short questionnaire assessing perceived synchronization, perceived transition dynamics, and overall immersion. A one-minute break separated the two rounds. The entire study lasted around 60 minutes.

\subsection{VR Scenario and Conditions}
We designed a VR mini-game in which participants cast magic from their right index finger to defeat scarecrow enemies. 
Red scarecrows required fire magic, whereas blue scarecrows required ice magic. 
Participants could carry only one magic type at a time. 
When carrying fire magic, they felt heat on the index fingertip; when carrying ice magic, they felt coolness at the same spot.

In each gameplay round, participants defeated six scarecrows. 
The color order was randomized while ensuring two blue-to-red transitions (Figure~\ref{fig:vrstudy}b-e), two red-to-blue transitions (Figure~\ref{fig:vrstudy}f-i), and one red-to-red sequence. The thermal stimuli corresponding to hot-to-cold and cold-to-hot transitions followed the same patterns used in Study~3.

The two conditions differed in how temperature was rendered. In the baseline condition, all four Peltier modules produced a uniform temperature, approximating a single actuator with the same total contact area. In the \thermosaic\ condition, one diagonal group began shifting toward the upcoming target temperature immediately after a scarecrow was defeated, while the other group remained within the current thermal category. This preserved the dominant percept while preparing a faster subsequent transition. When the participant switched magic type, all four modules were then set to the target thermal pattern.

\subsection{Results}
We analyzed both quantitative and qualitative data to evaluate how \thermosaic, compared to the baseline condition, affected perceived temporal alignment between thermal transitions and visual events in VR.

\textbf{Synchronization between thermal and visual changes.}
Participants were asked to rate the synchronization between the visual magic change (i.e., disappearance of the magic orb) and the perceived temperature change (Question 1, Q1) using a 0–100 visual analog scale (VAS). A score of 0 indicated complete asynchrony (i.e., a noticeable delay), whereas a score of 100 indicated near-perfect synchrony (i.e., occurring almost simultaneously).
The continuous nature of the VAS allowed participants to express fine-grained differences in perceived temporal alignment while reducing potential ceiling effects.

\begin{figure}[h]
    \centering
    \includegraphics[width=\columnwidth]{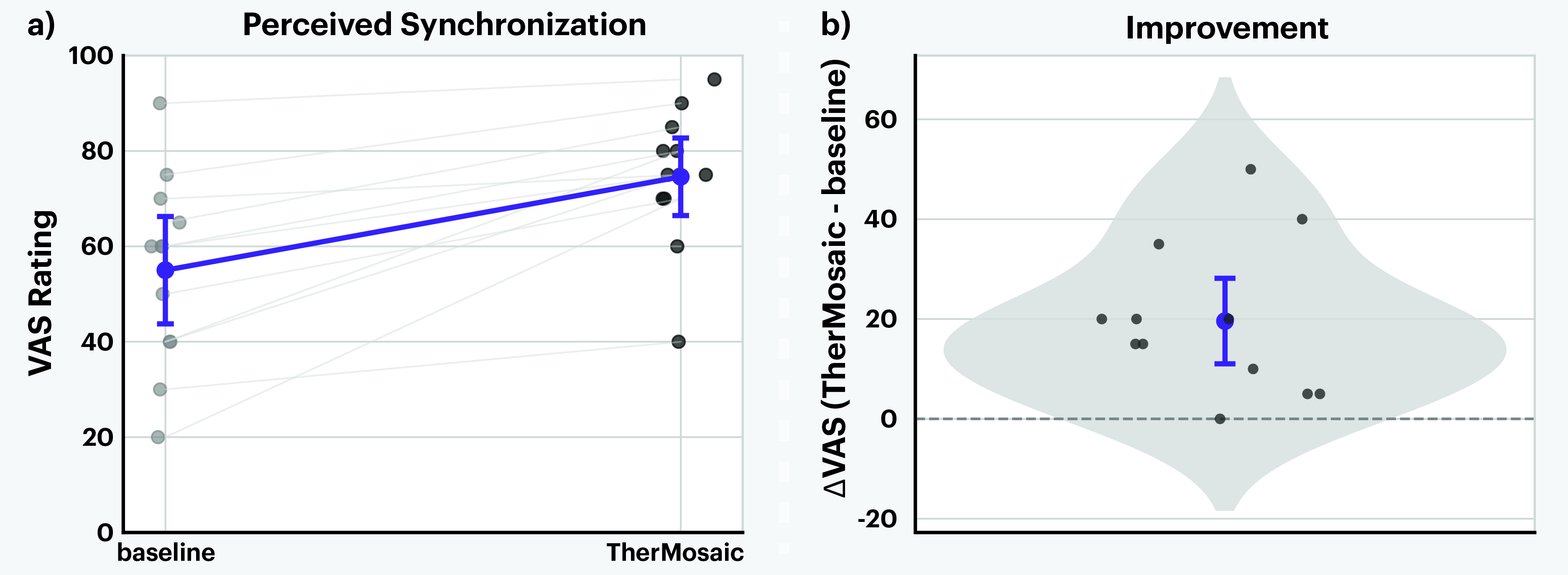}
    \caption{(a) VAS ratings for Q1 (“How synchronized did the magic switch and the temperature change feel?”) at baseline and under the \thermosaic condition. (b) Violin plot of individual improvements for $\Delta$VAS (\thermosaic{} $-$ baseline).}
    \Description{Panel (a) Participants’ VAS ratings Q1 (“How synchronized did the magic switch and the temperature change feel?”) at baseline and under the TherMosaic condition. Panel (b) Violin plot of individual improvements in synchronization (ΔVAS = TherMosaic − baseline), with most values above zero indicating improvement.}
    \label{fig:vr_vas}
\end{figure}

Participants reported significantly improved perceived synchronization under the \thermosaic condition compared to the baseline. 
As shown in Figure~\ref{fig:vr_vas}, VAS ratings increased by an average of 19.58 points (SD = 15.14), with a 95\% confidence interval of [11.01, 28.15]. 
A paired t-test confirmed this difference to be statistically significant ($t(11) = 4.48$, $p = .001$), with a large effect size (Cohen’s $d = 1.29$).

\begin{figure}[h]
    \centering
    \includegraphics[width=\columnwidth]{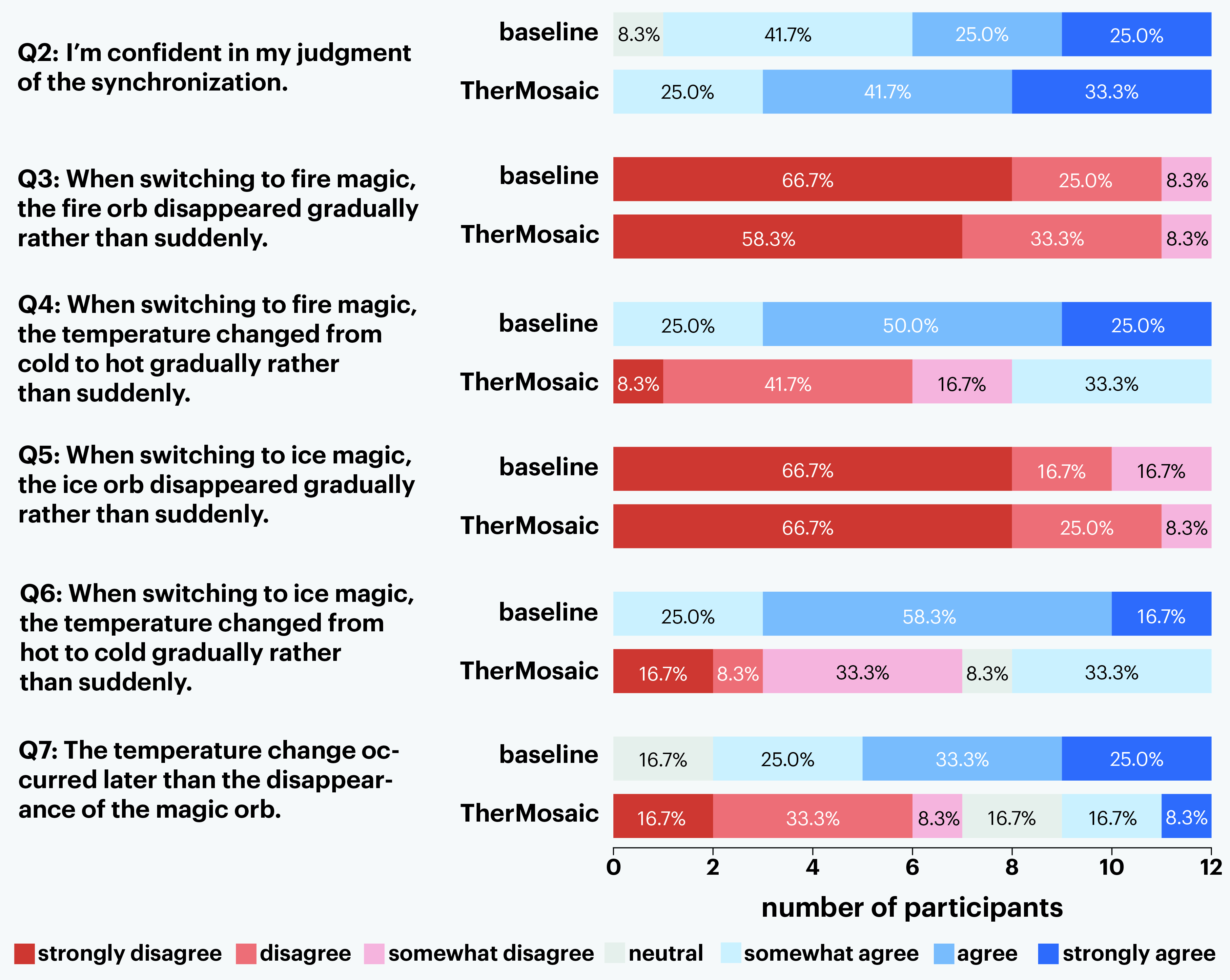}
    \caption{Participant Likert-scale responses for confidence (Q2), perceived visual abruptness (Q3, Q5), perceived thermal abruptness (Q4, Q6), and perceived thermal lag (Q7).}
    \Description{Stacked bar chart showing participant Likert-scale responses.}
    \label{fig:vr_results}
\end{figure}

Questions 2–7 are shown in Figure~\ref{fig:vr_results}. For Q2, participants reported high confidence in their judgments, with no significant difference between conditions ($\Delta = 0.42$, $p = 0.13$), suggesting that the observed improvement in synchrony perception is reliable and not driven by uncertainty or response noise.

\textbf{Modality-specific perception of transition dynamics.}
We examined how participants perceived the temporal dynamics of both visual and thermal changes across conditions.

For visual transitions, participants consistently rated the disappearance of the magic orb as sudden instead of gradual, with low mean ratings in both conditions (Q3, change to fire magic: $baseline = 1.42 \pm 0.19$, $\thermosaic = 1.50 \pm 0.19$; Q5, change to ice magic: $baseline = 1.50 \pm 0.23$, $\thermosaic = 1.42 \pm 0.19$), and no significant differences between conditions ($p > 0.5$). These results suggest that visual changes were perceived as nearly instantaneous across conditions.

In contrast, thermal perception differed substantially across conditions. Under the baseline condition, temperature changes were perceived as more gradual (Q4, change to fire magic: $6.00 \pm 0.21$; Q6, change to ice magic: $5.92 \pm 0.19$), whereas under \thermosaic, participants reported significantly more sudden transitions (Q4: $3.08 \pm 0.43$, $p < .001$; Q6: $3.33 \pm 0.43$, $p < .001$). 
This indicates that TherMosaic alters how temperature transitions are subjectively experienced, leading to a perception of more abrupt changes compared to the baseline.
This shift suggests a closer perceptual coupling between thermal changes and the rapid visual events.

Additionally, participants were significantly less likely to report that temperature changes lagged behind visual transitions under \thermosaic (Q7: $baseline = 5.67 \pm 0.31$, $\thermosaic = 3.17 \pm 0.53$, $p = .003$). 
This effect was consistent across both transition directions, suggesting that temperature changes are perceived as occurring more promptly under \thermosaic.

\textbf{Qualitative feedback.}
Qualitative feedback revealed a clear preference for the faster thermal transitions enabled by \thermosaic.
Participants described these transitions as more consistent with the visual feedback and less disruptive to the experience.
P2 noted that the slower baseline transitions felt ``weird'' because the perceived temperature no longer matched the visual state, while P12 stated that the faster changes ``match[ed] the visual magic better and feel[ed] more immersive.'' 
P3 further emphasized the functional impact of lag: 
\textit{``I do feel it affects me because if there is a delay, I’m not able to tell whether I caught the magic in my hand or not.''}
These comments further support the quantitative findings that \thermosaic\ improved temporal coherence between thermal and visual feedback.

\section{Discussion and Limitations}
Across three perceptual studies and a VR evaluation, our findings suggest that thermal latency can be addressed not only through improving hardware, but also through spatiotemporal thermal design. In the following, we interpret the perceptual mechanism suggested by these results, along with the practical limitations and future directions of \thermosaic.

\subsection{Interpreting the Perceptual Mechanism}
One plausible explanation for the observed effect is the interaction among several known properties of thermal perception. Thermal perception has relatively low spatial resolution, so distributed inputs across the fingertip may be integrated into a single dominant percept rather than being independently resolved. It is also often categorical, meaning that localized changes do not necessarily alter the perceived category immediately. In addition, thermal adaptation may reduce sensitivity to gradual local changes, allowing part of the stimulus to shift toward a target temperature without becoming salient. Together, these factors may allow local changes to remain perceptually masked within the dominant overall percept, reducing the remaining change required before the percept crosses into a new category. 

This interpretation is consistent with our three perceptual studies, which together support integration, perceptual stability under partial local change, and reduced transition time. While further work is needed to fully characterize the underlying mechanism, these results suggest that \thermosaic\ works by preconditioning part of the fingertip toward an upcoming target state before the full transition is consciously perceived.

\subsection{Limitations and Future Directions}
\subsubsection{Scheduling}
One limitation of \thermosaic is that for interactive use, it works best when upcoming thermal transitions can be anticipated. In our VR study, this was feasible because the interaction sequence explicitly controlled when temperature changes occurred. In more dynamic or spontaneous interactions, however, achieving the same effect may require a more sophisticated scheduling algorithm.

Future work could explore predictive control strategies that integrate user interaction modeling with real-time context awareness, as exemplified by scheduling approaches in~\cite{teng2022prolonging}. A similar framework for \thermosaic may anticipate upcoming temperature transitions and adapt feedback timing without breaking immersion.

\subsubsection{Scalability of Contact Area}
The current \thermosaic prototype targets a single fingertip, which was appropriate for isolating the perceptual mechanism but limits the range of interactions that can be supported.  In the VR study, P10 also noted that they expected more finger coverage and believed that involving a larger area would create a more solid sensation. 

Future work could extend thermal feedback to multiple fingers or full hands, enabling more immersive experiences such as grasping a warm object, feeling environmental coldness, or differentiating textures across fingers. 
Higher-density arrays may also support spatial gradients, thermal grill illusions, or thermal textures, though such scalability may introduce additional challenges in weight, cooling efficiency, and power management for a wearable device.

\subsubsection{Within-Category Intensity Invariance}
\thermosaic\ does not guarantee perfect within-category intensity invariance. Preconditioning may subtly alter perceived intensity within a category even when the categorical percept is preserved. 
Participants did not spontaneously report such changes in Study 3 or the VR study. 
Future work should directly quantify within-category intensity changes using continuous intensity ratings, just-noticeable-difference (JND) thresholds, or pairwise discrimination tasks.

\subsubsection{Cross-modal Perception}
\thermosaic\ shows that leveraging thermal perception alone can reduce perceived temperature transition time by up to 40\%. A promising next step is to combine thermal preconditioning with cross-modal cueing, such as visual or potentially auditory cues, to further strengthen this effect. For example, recent work by Günther et al.~\cite{10.1145/3613904.3642154} suggests that visual cues in VR can bias the perceived spatial location of physical thermal stimuli while still preserving authenticity, which implies that coordinated visual context may also help shape the perceived onset and coherence of thermal transitions in interactive environments.

\section{Conclusion}
We presented \thermosaic, a spatiotemporal thermal feedback approach for accelerating perceived thermal transitions on the fingertip by leveraging spatial summation and thermal adaptation.
Across three perceptual studies and a VR evaluation, we showed that heterogeneous thermal stimulation can reduce perceived thermal transition time and improve temporal coherence in interactive use.
These findings suggest that thermal latency is not solely a hardware limitation, but also an opportunity for perceptually informed interface design.

\begin{acks}
This work was supported in part by the National Science Foundation under Grant No. 2441833. During the preparation of this manuscript, the authors used LLM services to improve the clarity and linguistic flow of the text.
\end{acks}

\bibliographystyle{ACM-Reference-Format}
\bibliography{references}

\appendix

\section{Ramp-Rate Data}\label{app:ramp_data}
\begin{figure}[h]
    \centering
    \includegraphics[width=\columnwidth]{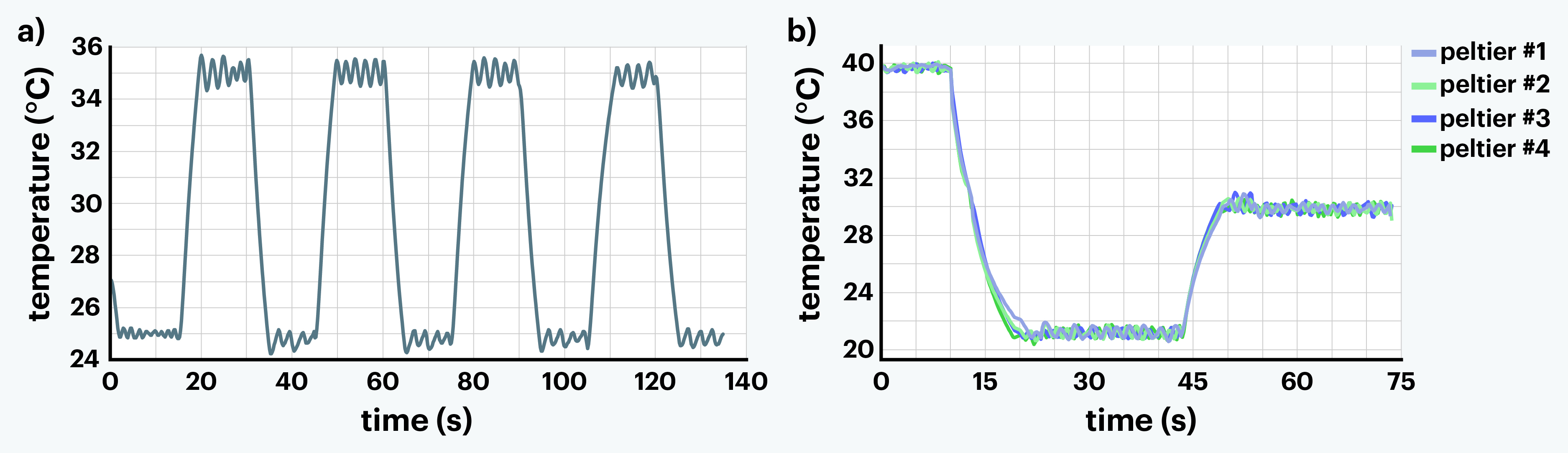}
    \caption{\thermosaic thermal performance. (a) Average step response between \SIrange{25}{35}{\celsius} across four Peltier modules. (b) Individual module responses during heating, cooling, and return to an intermediate setpoint, demonstrating stable convergence to target temperatures.}
    \Description{TherMosaic thermal performance. (a) The average step response between 25 and 35 °C across four Peltier modules. (b) Individual module responses during heating, cooling, and return to an intermediate setpoint, demonstrating stable convergence to the target temperatures.}
    \label{fig:ramp_avg}
\end{figure}

\textbf{Step response experiment.}
To characterize dynamic response, we applied repeated square-wave temperature setpoints alternating between \SI{25}{\celsius} and \SI{35}{\celsius}, as shown in Figure~\ref{fig:ramp_avg}a. 
Each step was held long enough for the system to reach steady state before switching. The plotted curve represents the average temperature across four Peltier modules at each timestamp.

Heating and cooling rates were computed from the transient segments using linear fits over the approximately linear portion of the temperature curve (typically between 10\%--90\% of the step amplitude). 
Each cycle produced one heating and one cooling estimate; reported values are averaged across multiple cycles.

\textbf{Multi-module transient behavior.}
To evaluate inter-module consistency, all four Peltier modules were driven simultaneously through a temperature sequence transitioning from a high (hot) setpoint to a low (cold) setpoint and then to an intermediate (neutral) level.
Temperatures from each module were recorded and plotted individually to capture variability across modules, as shown in Figure~\ref{fig:ramp_avg}b.

\begin{figure}[h]
    \centering
    \includegraphics[width=\columnwidth]{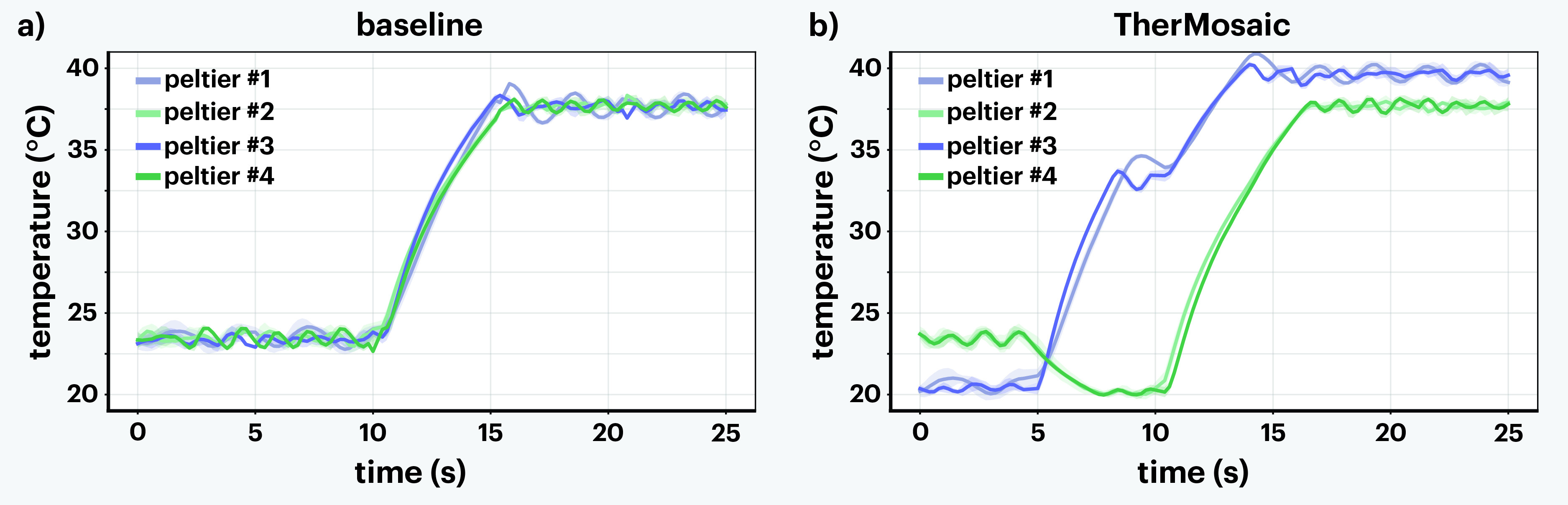}
    \caption{Temperature profiles of four Peltier modules during the cold-to-hot transition under (a) baseline and (b) \thermosaic conditions. Shaded areas denote SEM (n=5).}
    \Description{Line plots showing temperature profiles of four Peltier modules during a cold-to-hot transition under baseline and Thermosaic conditions. In the baseline condition, all modules change temperature at a similar rate. In the Thermosaic condition, modules change temperature in a staggered manner, with some modules heating earlier than others. Shaded regions indicate variability across trials (SEM, n = 5).}
    \label{fig:ramp_c2h}
\end{figure}

\begin{figure}[h]
    \centering
    \includegraphics[width=\columnwidth]{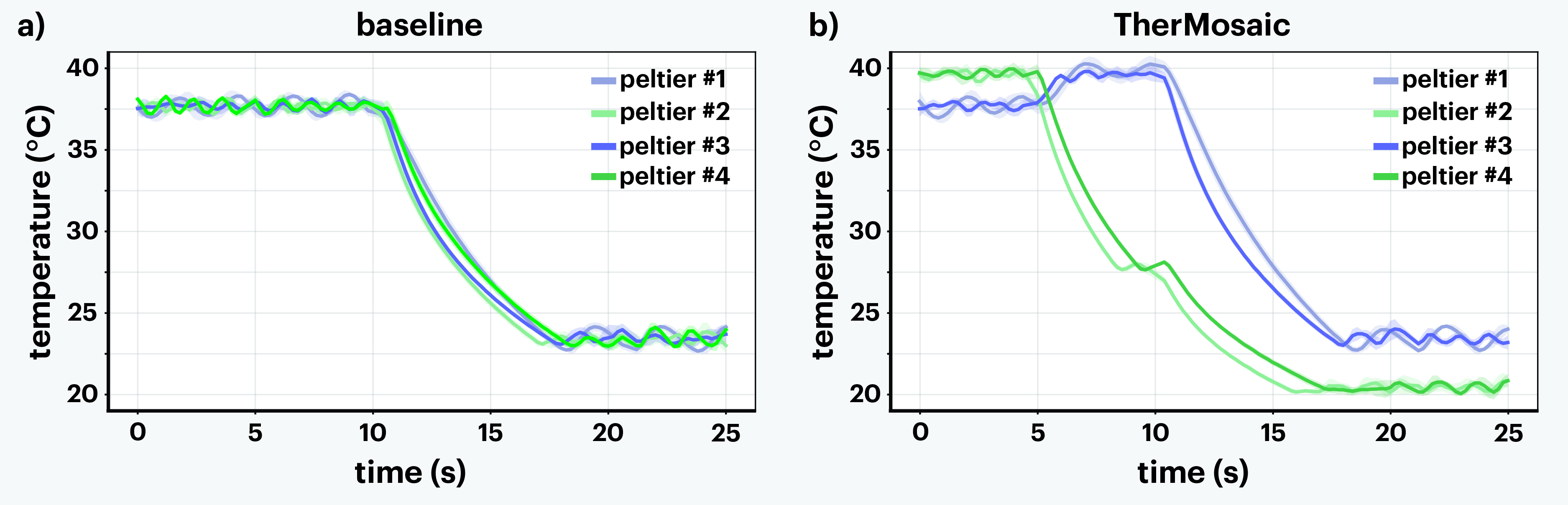}
    \caption{Temperature profiles of four Peltier modules during the hot-to-cold transition under (a) baseline and (b) \thermosaic conditions. Shaded areas denote SEM (n=5).}
    \Description{Line plots showing temperature profiles of four Peltier modules during a hot-to-cold transition under baseline and Thermosaic conditions. In the baseline condition, all modules change temperature at a similar rate. In the Thermosaic condition, modules change temperature in a staggered manner, with some modules cooling earlier than others. Shaded regions indicate variability across trials (SEM, n = 5).}
    \label{fig:ramp_h2c}
\end{figure}

\textbf{Representative cold-to-hot and hot-to-cold transitions.}
We measured the temporal temperature profiles of each Peltier module under representative transition settings, using the cold-to-hot and hot-to-cold transitions from Study 3 (Figure~\ref{fig:study3_result}a) as typical examples, with the ramp rate fixed at \SI{2}{\celsius\per\second} (Figure~\ref{fig:ramp_c2h} and Figure~\ref{fig:ramp_h2c}).

Each trial consisted of three sequential phases: Phase 1 (0--5~s) representing the initial steady state, Phase 2 (5--10~s) during which the \thermosaic preconditioning was triggered, and Phase 3 (10--25 s) capturing the target transition and steady state. 
These phase durations follow those used in Study 3 to ensure consistency, except that Phase 3 was extended by an additional 5 seconds to allow for more complete observation of the post-transition behavior.

We tested two conditions: baseline and \thermosaic. For each condition and transition type, the experiment was repeated five times.
The plotted curves represent the mean temperature across the five trials for each Peltier module, while the shaded regions indicate the standard error of the mean (SEM), reflecting variability across repetitions. 
This setup allows us to directly observe ramp rate and the spatial distribution of temperature over time.

\section{Thermal Calibration Procedure}
\label{app:thermal_cali}
To ensure reliable perception of cold, neutral, and hot categories for each participant, we calibrated four reference temperatures: cold anchor ($C$), cold--neutral boundary ($N_1$), neutral--hot boundary ($N_2$), and hot anchor ($H$). 
The procedure consisted of the following steps.

\textbf{Step 1: Initialization.}
Participants placed their fingertip on the thermal display at a neutral starting temperature (approximately \SI{30}{\celsius}) to familiarize themselves with the sensation. 
All stimuli were constrained within a safe operating range (\SIrange{20}{40}{\celsius}).

\textbf{Step 2: Estimation using discrete trials.}
Candidate temperatures near the expected category boundaries were presented in discrete trials. 
In each trial, the participant contacted the surface for 4\,s and reported whether the sensation was \textit{cold}, \textit{neutral}, or \textit{hot}. 
Based on the response, the researcher adjusted the next temperature in \SI{1}{\celsius} increments.

\textit{Cold--neutral boundary ($N_1$).}
Starting from lower temperatures, the stimulus was increased if the response was \textit{cold}, and decreased if the response was \textit{neutral}. This process was repeated until identifying the lowest temperature that was consistently judged as \textit{neutral}.

\textit{Cold anchor ($C$).}
Starting from much lower temperatures, the stimulus was increased if the response was \textit{cold}, and decreased if the response was \textit{neutral}. This process was repeated until identifying the highest temperature below $N_1$ that was consistently judged as \textit{cold} across repeated trials was selected as $C$.

\textit{Neutral--hot boundary ($N_2$).}
Starting from higher temperatures, the stimulus was decreased if the response was \textit{hot}, and increased if the response was \textit{neutral}. This process was repeated until identifying the highest temperature that was consistently judged as \textit{neutral}.

\textit{Hot anchor ($H$).}
Starting from much higher temperatures, the stimulus was decreased if the response was \textit{hot}, and increased if the response was \textit{neutral}. This process was repeated until identifying the lowest temperature above $N_2$ that was consistently judged as \textit{hot} across repeated trials was selected as $H$.

\textbf{Step 3: Verification.}
The final set $\{C, N_1, N_2, H\}$ was verified by presenting each temperature multiple times in random order. Participants were required to consistently classify each temperature according to its intended category before proceeding to the main experiment.

\end{document}